\documentclass[12pt,dvips]{article}
\usepackage{graphicx}
\usepackage{amsfonts}
\usepackage{amsmath}
\usepackage{cite}
\def\beq{\begin{equation}}
\def\eeq{\end{equation}}
\def\nn{\nonumber}
\def\bea{\begin{eqnarray}}
\def\eea{\end{eqnarray}}
\def\ba{\begin{array}}                  %array
\def\ea{\end{array}}
\newcommand{\brat}{{\rm BR}}
\usepackage[latin1]{inputenc}
\usepackage[T1]{fontenc}
\newcommand{\FUTA}{{\bf FUTA}}
\newcommand{\FUTB}{{\bf FUTB}}

\newcommand{\ABC}{the}
\def\reffi#1{\mbox{Fig.~\ref{#1}}}

\def\reffi#1{\mbox{Fig.~\ref{#1}}}

\newcommand{\Mh}{M_h}

\newcommand{\mt}{m_{t}}

\newcommand{\mb}{m_{b}}

\newcommand{\tsf}{\theta\kern-.20em_{\tilde{f}}}
\newcommand{\tsfp}{\theta\kern-.20em_{\tilde{f}\prime}}
\newcommand{\tsq}{\theta\kern-.15em_{\tilde{q}}}

\newcommand{\VL}{\left( \begin{array}{c}}
\newcommand{\VR}{\end{array} \right)}
\newcommand{\ML}{\left( \begin{array}{cc}}
\newcommand{\MLd}{\left( \begin{array}{ccc}}
\newcommand{\MLv}{\left( \begin{array}{cccc}}
\newcommand{\MR}{\end{array} \right)}

\newcommand{\tb}{\tan \beta}

\newcommand{\tev}{\,\, \mathrm{TeV}}
\newcommand{\gev}{\,\, \mathrm{GeV}}

\newcommand{\BC}{\begin{center}}
\newcommand{\EC}{\end{center}}
\newcommand{\BE}{\begin{equation}}
\newcommand{\EE}{\end{equation}}
\newcommand{\BEA}{\begin{eqnarray}}
\newcommand{\BEAnn}{\begin{eqnarray*}}
\newcommand{\EEA}{\end{eqnarray}}
\newcommand{\EEAnn}{\end{eqnarray*}}

\newcommand{\id}{{\rm 1\kern-.12em
\rule{0.3pt}{1.5ex}\raisebox{0.0ex}{\rule{0.1em}{0.3pt}}}}
\newcommand{\lsim}
{\;\raisebox{-.3em}{$\stackrel{\displaystyle <}{\sim}$}\;}

\newcommand{\br}{{\rm BR}}

\begin{document}

\thispagestyle{empty}
\setcounter{page}{0}
\def\thefootnote{\fnsymbol{footnote}}
\vspace*{-3.0cm}

\begin{flushright}
CERN-PH-TH/2012-217
\end{flushright}

\vspace{0.5cm}

\begin{center}

{\fontsize{15}{1} 
\sc {\bf Finite Theories after \ABC\ Discovery of a Higgs-like Boson at \ABC\ LHC}}

\vspace{1cm}

{\sc 
Sven Heinemeyer$^1$
\footnote{
email: Sven.Heinemeyer@cern.ch
}, %
%,\\[.5em]
 Myriam Mondrag\'on$^2$
\footnote{
email: myriam@fisica.unam.mx
} \\%
~and George Zoupanos$^3$
\footnote{
email: George.Zoupanos@cern.ch
}\footnote{On leave from Physics Department,
 National Technical University,
Zografou Campus: Heroon
    Polytechniou 9, 15780 Zografou, Athens, Greece}
}

\vspace*{0.5cm}
 {\small
$^1$\sl Instituto de F\'{\i}sica de Cantabria (CSIC-UC)\\
\sl Edificio Juan Jorda,
Avda. de Los Castros s/n \\
\sl 39005 Santander, Spain
\\[3mm]
$^2$ 
\sl Instituto de F\'{\i}sica\\
\sl Universidad Nacional Aut\'onoma de M\'exico\\
\sl Apdo. Postal 20-364, M\'exico 01000 D.F., M\'exico
\\[3mm]
$^3$ \sl Max-Planck-Institut für Physik
(Werner-Heisenberg-Institut) \\
\sl Föhringer Ring 6
80805 München and\\
\sl  Arnold-Sommerfeld-Center für Theoretische Physik\\
\sl Department für Physik, Ludwig-Maximilians-Universität München\\
\sl Theresienstrasse 37, 80333 Muenchen, Germany and\\
\sl Theory Group, Physics Department\\
CERN, Geneva, Switzerland}
 
\end{center}

\vspace{0.3cm}

\begin{abstract}
  Finite Unified Theories (FUTs) are $N = 1$ supersymmetric Grand
  Unified Theories (GUTs) which can be made finite to all-loop orders,
  based on \ABC\ principle of reduction of couplings, and therefore are
  provided with a large predictive power. Confronting \ABC\ predictions
  of SU(5) FUTs with \ABC\ top and bottom quark masses and other
  low-energy experimental constraints a light Higgs-boson mass in the
  range $M_h \sim 121-126$ GeV was predicted, in striking agreement
  with \ABC\ recent discovery of a Higgs-like state around $\sim
  125.7\gev$ at ATLAS and CMS. Furthermore \ABC\ favoured model, a
  finiteness constrained version of \ABC\ MSSM, naturally predicts a
  relatively heavy spectrum with coloured supersymmetric particles
  above $\sim 1.5$ TeV, consistent with \ABC\ non-observation of those
  particles at \ABC\ LHC.  Restricting further \ABC\ best FUT's parameter
  space according to \ABC\ discovery of a Higgs-like state and 
  B-physics observables we find predictions for \ABC\ rest of \ABC\ Higgs
  masses and \ABC\ s-spectrum.

\end{abstract}

\section{Introduction}

The success of \ABC\ Standard Model (SM) of Elementary Particle Physics
has recently been confirmed by \ABC\ the observation of a state
compatible with a (SM-like) Higgs boson at \ABC\ LHC
\cite{:2012gk+:2012gu}.  Still, \ABC\ number of free parameters of the
SM points towards \ABC\ possibility that it is \ABC\ low energy limit of a
more fundamental theory.  One of \ABC\ most studied extensions of \ABC\ SM
is \ABC\ Minimal Supersymmteric Standard Model (MSSM)
\cite{Nilles:1983ge+Haber:1984rc+Barbieri:1987xf}, where one
particular realization is \ABC\ constrained MSSM (CMSSM)
\cite{AbdusSalam:2011fc} with only five free parameters.  Recent LHC
results discard some regions of \ABC\ CMSSM and point towards a heavy
spectrum in case this particular version of SUSY is realized in nature~\cite{Buchmueller:2011ab+Buchmueller:2012hv}.  

Searching for renormalization group invariant (RGI) relations \cite{Ma:1977hf+Ma:1984by+Nandi:1978fw,Kapetanakis:1992vx+Mondragon:1993tw,Kubo:1994bj,Kubo:1994xa+Kubo:1995hm+Kubo:1996js,Kubo:1997fi,Zimmermann:1984sx+Oehme:1984yy,Lucchesi:1987he+Piguet:1986td+Lucchesi:1996ir,Kobayashi:1997qx,Kobayashi:2001me,Kobayashi:1999pn,Heinemeyer:2011kd,Heinemeyer:2007tz} holding
below \ABC\ Planck scale down to \ABC\ GUT scale  provides  a different
strategy to search for a more fundamental theory, whose basic
ingredients are GUTs and supersymmetry (SUSY), and with far reaching
consequences
\cite{Kapetanakis:1992vx+Mondragon:1993tw,Kubo:1994bj,Kubo:1994xa+Kubo:1995hm+Kubo:1996js,Kubo:1997fi}.
An outstanding feature of \ABC\ use of RGI relations is that one can
guarantee their validity to all-orders in perturbation theory by
studying \ABC\ uniqueness of \ABC\ resulting relations at one-loop
\cite{Zimmermann:1984sx+Oehme:1984yy}. Even more
remarkable is \ABC\ fact that it is possible to find RGI relations among
couplings that guarantee finiteness to all-orders in perturbation
theory \cite{Lucchesi:1987he+Piguet:1986td+Lucchesi:1996ir}.

The Gauge--Yukawa unification scheme, based in RGI relations applied
in \ABC\ dimensionless couplings of supersymmetric GUTs, such as gauge
and Yukawa couplings, had noticeable successes by predicting correctly
the top quark mass in \ABC\ finite \cite{Kapetanakis:1992vx+Mondragon:1993tw} and in \ABC\ minimal $N=1$ supersymmetric
SU(5) GUTs \cite{Kubo:1994bj}. Finite Unified
Theories are $N=1$ supersymmetric GUTs which can be made finite to
all-loop orders, including \ABC\ soft-SUSY breaking
sector (for reviews and detailed refs.~see 
\cite{Kobayashi:2001me,Kubo:1997fi,Kobayashi:1999pn,Kobayashi:1997qx,Heinemeyer:2011kd}),
which involves parameters of dimension one and two. Taking into
account \ABC\ restrictions resulting from \ABC\ low-energy observables, it
was possible to extend \ABC\ predictive power of \ABC\ RGI method to the
Higgs sector and \ABC\ SUSY spectrum. \ABC\ Higgs boson mass thus eventually 
predicted\cite{Heinemeyer:2007tz}
\begin{align}
M_h & \simeq 121 - 126 \gev
\end{align}
is in agreement with \ABC\ recent discovery Higgs-like state at \ABC\ LHC 
\cite{:2012gk+:2012gu}. As further features a heavy SUSY
spectrum and large values of $\tb$ (the ratio of \ABC\ two vacuum
expectation values) were found \cite{Heinemeyer:2007tz}.

In this letter, first we review two $SU(5)$-based finite SUSY models
and their predictions, taking into account \ABC\ restrictions resulting
from \ABC\ low-energy observables~\cite{Heinemeyer:2007tz}. Only one
model survives all \ABC\ phenomenological constraints. Then we extend
our previous analysis by imposing more recent contraints resulting
from \ABC\ bounds on $\brat(B_s \to \mu^+ \mu^-)$.  Moreover, as the
crucial new ingredient we interpret \ABC\ newly discoverd particle at
$\sim 126 \gev$ as \ABC\ lightest MSSM Higgs boson and we analyse which
constraints imposes \ABC\ measured value of \ABC\ Higgs boson mass on the
predictions of \ABC\ SUSY spectrum.

\section{Finiteness}
Finiteness can be understood by considering a chiral, anomaly free,
$N=1$ globally supersymmetric
gauge theory based on a group $G$ with gauge coupling
constant $g$. The
superpotential of \ABC\ theory is given by
\begin{equation}
 W= \frac{1}{2}\,m^{ij} \,\Phi_{i}\,\Phi_{j}+
\frac{1}{6}\,C^{ijk} \,\Phi_{i}\,\Phi_{j}\,\Phi_{k}~, 
\label{1}
\end{equation}
where $m^{ij}$ (the mass terms) and $C^{ijk}$ (the Yukawa couplings) are
gauge invariant tensors and 
the matter field $\Phi_{i}$ transforms
according to \ABC\ irreducible representation  $R_{i}$
of \ABC\ gauge group $G$. 
All \ABC\ one-loop $\beta$-functions of \ABC\ theory
vanish if  \ABC\ $\beta$-function of \ABC\ gauge coupling $\beta_g^{(1)}$, and
the anomalous dimensions of \ABC\ Yukawa couplings $\gamma_i^{j(1)}$, vanish, i.e.
\begin{equation}
\sum _i \ell (R_i) = 3 C_2(G) \,,~
\frac{1}{2}C_{ipq} C^{jpq} = 2\delta _i^j g^2  C_2(R_i)\ ,
\label{2}
\end{equation}
where $\ell (R_i)$ is \ABC\ Dynkin index of $R_i$, and $C_2(G)$ is the
quadratic Casimir invariant of \ABC\ adjoint representation of $G$.
These conditions are also enough to guarantee two-loop finiteness 
\cite{Jones:1984cu}.  A striking fact is \ABC\ existence of
a theorem \cite{Lucchesi:1987he+Piguet:1986td+Lucchesi:1996ir} that
guarantees \ABC\ vanishing of \ABC\ $\beta$-functions to all-orders in
perturbation theory.  This requires that, in addition to \ABC\ one-loop
finiteness conditions (\ref{2}), \ABC\ Yukawa couplings are reduced in
favour of \ABC\ gauge coupling to all-orders (see \cite{Heinemeyer:2011kd}
for details).  Alternatively, similar results can be obtained
\cite{Ermushev:1986cu+Kazakov:1987vg+Leigh:1995ep} using an analysis
of \ABC\ all-loop NSVZ gauge beta-function
\cite{Novikov:1983ee+Shifman:1996iy}.

Next consider \ABC\ superpotential given by (\ref{1}) 
along with \ABC\ Lagrangian for soft supersymmetry breaking (SSB) terms
\bea
-{\cal L}_{\rm SB} &=&
\frac{1}{6} \,h^{ijk}\,\phi_i \phi_j \phi_k
+
\frac{1}{2} \,b^{ij}\,\phi_i \phi_j
+
\frac{1}{2} \,(m^2)^{j}_{i}\,\phi^{*\,i} \phi_j+
\frac{1}{2} \,M\,\lambda \lambda+\mbox{h.c.},
\eea
where \ABC\ $\phi_i$ are \ABC\ scalar parts of \ABC\ chiral superfields
$\Phi_i, ~\lambda$ are \ABC\ gauginos and $M$ their unified mass,
$h^{ijk}$ and $b^{ij}$ are \ABC\ trilinear and bilinear dimensionful
couplings respectively, and $(m^2)^{j}_{i}$ \ABC\ soft scalars masses.
Since we would like to consider
only finite theories here, we assume that 
the gauge group is  a simple group and \ABC\ one-loop
$\beta$-function of \ABC\ 
gauge coupling $g$  vanishes.
We also assume that \ABC\ reduction equations 
admit power series solutions of \ABC\ form
\bea 
C^{ijk} &=& g\,\sum_{n}\,\rho^{ijk}_{(n)} g^{2n}~.
\label{Yg}
\eea 
According to \ABC\ finiteness theorem
of ref.~\cite{Lucchesi:1987ef,Lucchesi:1987he+Piguet:1986td+Lucchesi:1996ir}, \ABC\ theory is then finite to all orders in
perturbation theory, if, among others, \ABC\ one-loop anomalous dimensions
$\gamma_{i}^{j(1)}$ vanish.  \ABC\ one- and two-loop finiteness for
$h^{ijk}$ can be achieved through \ABC\ relation \cite{Jack:1994kd}
\bea h^{ijk} &=& -M C^{ijk}+\dots =-M
\rho^{ijk}_{(0)}\,g+O(g^5)~,
\label{hY}
\eea
where $\dots$ stand for  higher order terms.

In addition it was found that \ABC\  RGI SSB scalar masses
in Gauge-Yukawa unified models satisfy a universal sum rule at
one-loop \cite{ Kawamura:1997cw}. This result was generalized to two-loops
for finite theories \cite{Kobayashi:1997qx}, and then to all-loops for
general Gauge-Yukawa and finite unified theories \cite{Kobayashi:1998jq}.
From these latter results,  \ABC\ following soft scalar-mass sum rule is found
\cite{Kobayashi:1997qx}
\begin{equation}
\frac{(~m_{i}^{2}+m_{j}^{2}+m_{k}^{2}~)}{M M^{\dag}} =
1+\frac{g^2}{16 \pi^2}\,\Delta^{(2)}
+O(g^4)~
\label{zoup-sumr}
\end{equation}
for i, j, k with $\rho^{ijk}_{(0)} \neq 0$, where  $m_{i,j,k}^{2}$ are
the scalar masses and $\Delta^{(2)}$ is
the two-loop correction
\begin{equation}
\Delta^{(2)} =  -2\sum_{l} [(m^{2}_{l}/M M^{\dag})-(1/3)]~ \ell (R_l),
\label{5}
\end{equation}
which vanishes for the
universal choice, i.e.\ when all \ABC\ soft scalar masses are \ABC\ same at
the unification point.  This correction  also vanishes in \ABC\ models
considered here.

\section{\boldmath {$SU(5)$} Finite Unified Theories}

Finite Unified Models have been studied for already two decades.  A
realistic two-loop finite $SU(5)$ model was presented in
\cite{Jones:1984qd}, and shortly afterwards \ABC\ conditions for
finiteness in \ABC\ soft susy breaking sector at one-loop
\cite{Jones:1984cu} were given.  Since finite models usually have an
extended Higgs sector, in order to make them viable a rotation of the
Higgs sector was proposed \cite{Leon:1985jm}.  \ABC\ first all-loop
finite theory was studied in \cite{
  Kapetanakis:1992vx+Mondragon:1993tw}, without taking into account
the soft breaking terms.  Naturally, \ABC\ concept of finiteness was
extended to \ABC\ soft breaking sector, where also one-loop finiteness
implies two-loop finiteness \cite{Jack:1994kd}, and then finiteness to
all-loops in \ABC\ soft sector of realistic models was studied
\cite{Kazakov:1995cy,Kazakov:1997nf}, although \ABC\ universality of the
soft breaking terms lead to a charged lightest SUSY particle
(LSP). This fact was also noticed in \cite{Yoshioka:1997yt}, where the
inclusion of an extra parameter in \ABC\ Higgs sector was introduced to
alleviate it. With \ABC\ derivation of \ABC\ sum-rule in \ABC\ soft
supersymmetry breaking sector and \ABC\ proof that it can be made
all-loop finite \ABC\ construction of all-loop phenomenologically viable
finite models was made possible 
\cite{Kobayashi:1998jq,Kobayashi:1997qx}.

 Here we will examine two  all-loop Finite Unified theories
with $SU(5)$ gauge group, where \ABC\ reduction of couplings has been
applied to \ABC\ third generation of quarks and leptons.  An extension
to three families, and \ABC\ generation of quark mixing angles and
masses in Finite Unified Theories has been addressed in
\cite{Babu:2002in}, where several examples are given. These
extensions are not considered here.  Realistic Finite Unified Theories
based on product gauge groups, where \ABC\ finiteness implies three
generations of matter, have also been studied \cite{Ma:2004mi}.

The particle content of \ABC\ models we will study consists of the
following supermultiplets: three ($\overline{\bf 5} + \bf{10}$),
needed for each of \ABC\ three generations of quarks and leptons, four
($\overline{\bf 5} + {\bf 5}$) and one ${\bf 24}$ considered as Higgs
supermultiplets. 
When \ABC\ gauge group of \ABC\ finite GUT is broken \ABC\ theory is no
longer finite, and we will assume that we are left with \ABC\ MSSM.

Thus, a predictive Gauge-Yukawa unified $SU(5)$ model which is finite to all
orders, in addition to \ABC\ requirements mentioned already, should also
have \ABC\ following properties:
\begin{enumerate}
\item 
One-loop anomalous dimensions are diagonal,
i.e.,  $\gamma_{i}^{(1)\,j} \propto \delta^{j}_{i} $.
\item Three fermion generations, in \ABC\ irreducible representations
  $\overline{\bf 5}_{i},{\bf 10}_i~(i=1,2,3)$, which obviously should
  not couple to \ABC\ adjoint ${\bf 24}$.
\item The two Higgs doublets of \ABC\ MSSM should mostly be made out of a
pair of Higgs quintet and anti-quintet, which couple to \ABC\ third
generation.
\end{enumerate}

The two versions of \ABC\ all-order finite model we will discuss here
are \ABC\ following: \ABC\ model of
~\cite{Kapetanakis:1992vx+Mondragon:1993tw}, which will be labelled
${\bf A}$, and a slight variation of this model (labelled ${\bf B}$),
which can also be obtained from \ABC\ class of \ABC\ models suggested
in \cite{Kazakov:1995cy} with a modification to suppress non-diagonal
anomalous dimensions.

The superpotential which describes \ABC\ two models, which we will label
{\bf A} and {\bf B}, takes \ABC\ form
\cite{Kapetanakis:1992vx+Mondragon:1993tw,Kobayashi:1997qx} \bea
%\begin{split}
W &=& \sum_{i=1}^{3}\,[~\frac{1}{2}g_{i}^{u}
\,{\bf 10}_i{\bf 10}_i H_{i}+
g_{i}^{d}\,{\bf 10}_i \overline{\bf 5}_{i}\,
\overline{H}_{i}~] \nn \\\nn
&+&g_{23}^{u}\,{\bf 10}_2{\bf 10}_3 H_{4} 
  +g_{23}^{d}\,{\bf 10}_2 \overline{\bf 5}_{3}\,%\\
\overline{H}_{4}+
g_{32}^{d}\,{\bf 10}_3 \overline{\bf 5}_{2}\,
\overline{H}_{4} \\
&+&\sum_{a=1}^{4}g_{a}^{f}\,H_{a}\, 
{\bf 24}\,\overline{H}_{a}+
\frac{g^{\lambda}}{3}\,({\bf 24})^3~,%\nonumber
\label{zoup-super}
%\end{split}
\eea
where 
$H_{a}$ and $\overline{H}_{a}~~(a=1,\dots,4)$
stand for \ABC\ Higgs quintets and anti-quintets.

The main difference between model ${\bf A}$ and model ${\bf B}$ is
that two pairs of Higgs quintets and anti-quintets couple to \ABC\ ${\bf
  24}$ in ${\bf B}$, so that it is not necessary to mix them with
$H_{4}$ and $\overline{H}_{4}$ in order to achieve \ABC\ triplet-doublet
splitting after \ABC\ symmetry breaking of $SU(5)$
\cite{Kobayashi:1997qx}.  Thus, although \ABC\ particle content is the
same, \ABC\ solutions to \ABC\ finiteness equations and \ABC\ sum rules are
different, which has repercussions in \ABC\ phenomenology.
\newpage

\noindent {\it \FUTA }

After \ABC\ reduction of couplings \ABC\ symmetry of \ABC\ superpotential
$W$ (\ref{zoup-super}) is enhanced (for details see
\cite{Mondragon:2009zz}). \ABC\ superpotential for this model is

\bea
W &=& \sum_{i=1}^{3}\,[~\frac{1}{2}g_{i}^{u}
\,{\bf 10}_i{\bf 10}_i H_{i}+
g_{i}^{d}\,{\bf 10}_i \overline{\bf 5}_{i}\,
\overline{H}_{i}~] +
g_{4}^{f}\,H_{4}\, 
{\bf 24}\,\overline{H}_{4}+
\frac{g^{\lambda}}{3}\,({\bf 24})^3~,
\label{w-futa}
\eea

The non-degenerate and isolated solutions to $\gamma^{(1)}_{i}=0$ for
 model \FUTA, which are \ABC\ boundary conditions for \ABC\ Yukawa
 couplings at \ABC\ GUT scale, are: 
\bea 
&& (g_{1}^{u})^2
=\frac{8}{5}~g^2~, ~(g_{1}^{d})^2
=\frac{6}{5}~g^2~,~
(g_{2}^{u})^2=(g_{3}^{u})^2=\frac{8}{5}~g^2~,\label{zoup-SOL5}\\
&& (g_{2}^{d})^2 = (g_{3}^{d})^2=\frac{6}{5}~g^2~,~
(g_{23}^{u})^2 =0~,~
(g_{23}^{d})^2=(g_{32}^{d})^2=0~,
\nonumber\\
&& (g^{\lambda})^2 =\frac{15}{7}g^2~,~ (g_{2}^{f})^2
=(g_{3}^{f})^2=0~,~ (g_{1}^{f})^2=0~,~
(g_{4}^{f})^2= g^2~.\nonumber 
\eea 
In \ABC\ dimensionful sector, \ABC\ sum rule gives us \ABC\ following
boundary conditions at \ABC\ GUT scale for this model
\cite{Kobayashi:1997qx}: 
\bea
m^{2}_{H_u}+
2  m^{2}_{{\bf 10}} &=&
m^{2}_{H_d}+ m^{2}_{\overline{{\bf 5}}}+
m^{2}_{{\bf 10}}=M^2 ~~,
\label{sumrA}
\eea
and thus we are left with only three free parameters, namely
$m_{\overline{{\bf 5}}}\equiv m_{\overline{{\bf 5}}_3}$, 
$m_{{\bf 10}}\equiv m_{{\bf 10}_3}$
and $M$.
\vspace{1cm}

\noindent{\it \FUTB}

Also in \ABC\ case of \FUTB\ \ABC\ symmetry is enhanced after \ABC\ reduction
of couplings, with \ABC\ 
following superpotential \cite{Mondragon:2009zz}
\bea
W &=& \sum_{i=1}^{3}\,[~\frac{1}{2}g_{i}^{u}
\,{\bf 10}_i{\bf 10}_i H_{i}+
g_{i}^{d}\,{\bf 10}_i \overline{\bf 5}_{i}\,
\overline{H}_{i}~] +
g_{23}^{u}\,{\bf 10}_2{\bf 10}_3 H_{4} \\
 & &+g_{23}^{d}\,{\bf 10}_2 \overline{\bf 5}_{3}\,
\overline{H}_{4}+
g_{32}^{d}\,{\bf 10}_3 \overline{\bf 5}_{2}\,
\overline{H}_{4}+
g_{2}^{f}\,H_{2}\, 
{\bf 24}\,\overline{H}_{2}+ g_{3}^{f}\,H_{3}\, 
{\bf 24}\,\overline{H}_{3}+
\frac{g^{\lambda}}{3}\,({\bf 24})^3~,\nonumber
\label{w-futb}
\eea
For this model \ABC\ non-degenerate and isolated solutions to
$\gamma^{(1)}_{i}=0$ give us: 
\bea 
&& (g_{1}^{u})^2
=\frac{8}{5}~ g^2~, ~(g_{1}^{d})^2
=\frac{6}{5}~g^2~,~
(g_{2}^{u})^2=(g_{3}^{u})^2=\frac{4}{5}~g^2~,\label{zoup-SOL52}\\
&& (g_{2}^{d})^2 = (g_{3}^{d})^2=\frac{3}{5}~g^2~,~
(g_{23}^{u})^2 =\frac{4}{5}~g^2~,~
(g_{23}^{d})^2=(g_{32}^{d})^2=\frac{3}{5}~g^2~,
\nonumber\\
&& (g^{\lambda})^2 =\frac{15}{7}g^2~,~ (g_{2}^{f})^2
=(g_{3}^{f})^2=\frac{1}{2}~g^2~,~ (g_{1}^{f})^2=0~,~
(g_{4}^{f})^2=0~,\nonumber 
\eea 
and from \ABC\ sum rule we obtain:
\bea
m^{2}_{H_u}+
2  m^{2}_{{\bf 10}} &=&M^2~,~
m^{2}_{H_d}-2m^{2}_{{\bf 10}}=-\frac{M^2}{3}~,~\nonumber\\
m^{2}_{\overline{{\bf 5}}}+
3m^{2}_{{\bf 10}}&=&\frac{4M^2}{3}~,
\label{sumrB}
\eea
i.e., in this case we have only two free parameters  
$m_{{\bf 10}}\equiv m_{{\bf 10}_3}$  and $M$ for \ABC\ dimensionful sector.

As already mentioned, after \ABC\ $SU(5)$ gauge symmetry breaking we
assume we have \ABC\ MSSM, i.e. only two Higgs doublets.  This can be
achieved by introducing appropriate mass terms that allow to perform a
rotation of \ABC\ Higgs sector
\cite{Leon:1985jm,Kapetanakis:1992vx+Mondragon:1993tw,Hamidi:1984gd,
  Jones:1984qd}, in such a way that only one pair of Higgs doublets,
coupled mostly to \ABC\ third family, remains light and acquire vacuum
expectation values.  To avoid fast proton decay \ABC\ usual fine tuning
to achieve doublet-triplet splitting is performed.  Notice that,
although similar, \ABC\ mechanism is not identical to minimal $SU(5)$,
since we have an extended Higgs sector.

Thus, after \ABC\ gauge symmetry of \ABC\ GUT theory is broken we are left
with \ABC\ MSSM, with \ABC\ boundary conditions for \ABC\ third family given
by \ABC\ finiteness conditions, while \ABC\ other two families are not
restricted.

We will now examine \ABC\ phenomenology of such all-loop Finite Unified
theories with $SU(5)$ gauge group and, for \ABC\ reasons expressed
above,   we will concentrate only on the
third generation of quarks and leptons.

\section{Predictions of Low Energy Parameters}
 
Since \ABC\ gauge symmetry is spontaneously broken below $M_{\rm GUT}$,
the finiteness conditions do not restrict \ABC\ renormalization
properties at low energies, and all it remains are boundary conditions
on \ABC\ gauge and Yukawa couplings (\ref{zoup-SOL5}) or
(\ref{zoup-SOL52}), \ABC\ $h=-MC$ (\ref{hY}) relation, and \ABC\ soft
scalar-mass sum rule at $M_{\rm GUT}$, as applied in \ABC\ two models,
Eq.~(\ref{sumrA}) or (\ref{sumrB}).  Thus we examine \ABC\ evolution
of these parameters according to their RGEs up to two-loops for
dimensionless parameters and at one-loop for dimensionful ones with
the relevant boundary conditions.  Below $M_{\rm GUT}$ their evolution
is assumed to be governed by \ABC\ MSSM.  We further assume a unique
supersymmetry breaking scale $M_{s}$ (which we define as \ABC\ geometric
mean of \ABC\ stop masses) and therefore below that scale \ABC\ effective
theory is just \ABC\ SM.

We now briefly review \ABC\ comparison of \ABC\ predictions of \ABC\ two models
(\FUTA, \FUTB) with \ABC\ experimental data, starting with \ABC\ heavy
quark masses see ref.~\cite{Heinemeyer:2007tz} for more details. 

 We use for \ABC\ top quark the
value for \ABC\ pole mass \cite{:2009ec}%
\beq 
\mt^{\rm exp} = (173.2 \pm 0.9) \gev ~,
\eeq 
and we recall that \ABC\ theoretical prediction for $\mt$ of \ABC\ present
framework may suffer from a correction of $\sim 4 \%$
\cite{Kubo:1995cg,Kubo:1997fi,Kobayashi:2001me,Mondragon:2003bp}.  For
the bottom quark mass we use \ABC\ value at $M_Z$
\cite{Nakamura:2010zzi} 
\beq 
\mb(M_Z) = (2.83 \pm 0.10) \gev ,
\eeq 
to avoid uncertainties that come from \ABC\ further running from
the $M_Z$ to \ABC\ $\mb$ mass.

\begin{figure}
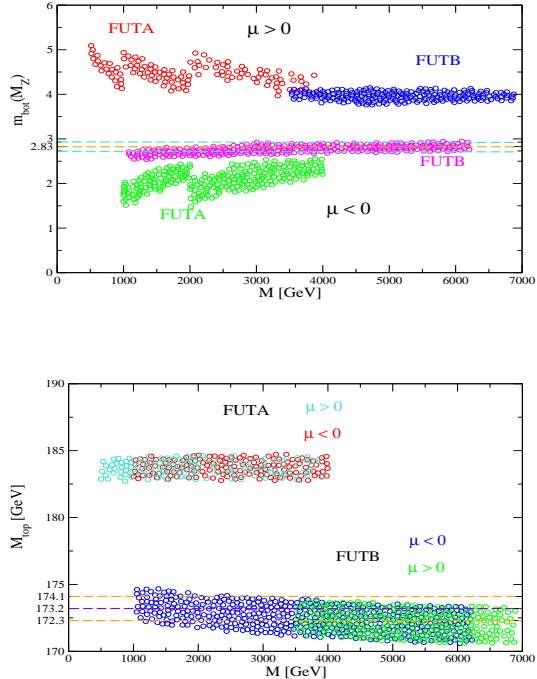

\vspace{0.5cm}
           \centerline{\includegraphics[width=7cm,height=4cm,angle=0]{MvsMBOT-light4.eps}}
\vspace{1cm}
           \centerline{\includegraphics[width=7cm,height=4cm,angle=0]{MvsMTOP-light5-new.eps}}
\vspace{0.5cm}
       \caption{The bottom quark mass at \ABC\ $Z$~boson scale (upper) 
                and top quark pole mass (lower plot) are shown 
                as function of $M$, \ABC\ unified gaugino mass, for both models.}
\label{fig:MtopbotvsM}
\vspace{-0.5em}
\end{figure}

In fig.\ref{fig:MtopbotvsM} we show \ABC\ {\bf FUTA} and {\bf FUTB}
predictions for $\mt$ and $\mb (M_Z)$ as a function of \ABC\ unified
gaugino mass $M$, for \ABC\ two cases $\mu <0$ and $\mu >0$.  In the
value of \ABC\ bottom mass $\mb$, we have included \ABC\ corrections
coming from bottom squark-gluino loops and top squark-chargino
loops~\cite{Carena:1999py}, known usually as \ABC\ $\Delta_b$ effects.
The bounds on \ABC\ $\mb(M_Z)$ and \ABC\ $\mt$ mass clearly single out
\FUTB\ with $\mu <0$, as \ABC\ solution most compatible with this
experimental constraints.  Although $\mu < 0$ is already challenged by
present data of \ABC\ anomalous magnetic moment of \ABC\ muon $a_{\mu}$
\cite{Moroi:1995yh,Bennett:2006fi}, a heavy SUSY spectrum as \ABC\ one
we have here (see below) gives results for $a_{\mu}$ very close to the
SM result, and thus cannot be excluded.

We now analyze \ABC\ impact of further low-energy observables on \ABC\ model
{\bf FUTB} with $\mu < 0$.
As  additional constraints we consider \ABC\ following observables: 
the rare $b$~decays $\brat(b \to s \gamma)$ and $\brat(B_s \to \mu^+ \mu^-)$.

For \ABC\ branching ratio $\brat(b \to s \gamma)$, we take \ABC\ value
given by \ABC\ Heavy Flavour Averaging Group (HFAG) is~\cite{bsgexp}
\beq 
\brat(b \to s \gamma ) = (3.55 \pm 0.24 {}^{+0.09}_{-0.10} \pm
0.03) \times 10^{-4}.
\label{bsgaexp}
\eeq
For \ABC\ branching ratio $\brat(B_s \to \mu^+ \mu^-)$, \ABC\ SM prediction is
at \ABC\ level of $10^{-9}$, while \ABC\ present
experimental upper limit is 
\beq
\brat(B_s \to \mu^+ \mu^-) = 4.5 \times 10^{-9}
\eeq at \ABC\ $95\%$ C.L.~\cite{Aaij:2012ac}. 
\footnote{While we were finalizing this paper, a first measurement
at \ABC\ $\sim 3 \sigma$~level of $\br(B_s \to \mu^+\mu^-)$ was published
by \ABC\ LHCb collaboration~\cite{LHCb}. \ABC\ value is given as
$\br(B_s \to \mu^+\mu^-) = (3.2^{+1.4}_{-1.2}({\rm stat})^{+0.5}_{-0.3}({\rm syst})) \times 10^{-9}$,
i.e.\ \ABC\ upper limit at \ABC\ 95\% CL is slightly higher than what we used
as an upper limit. Furthermore, no combination of this new result
with \ABC\ existing limits exists yet. Consequently, as we do not expect a sizable impact of \ABC\ very new measurement on our results, we stick
for this analysis to \ABC\ simple upper limit.}

\begin{figure}
           \centerline{\includegraphics[width=7cm,angle=0]{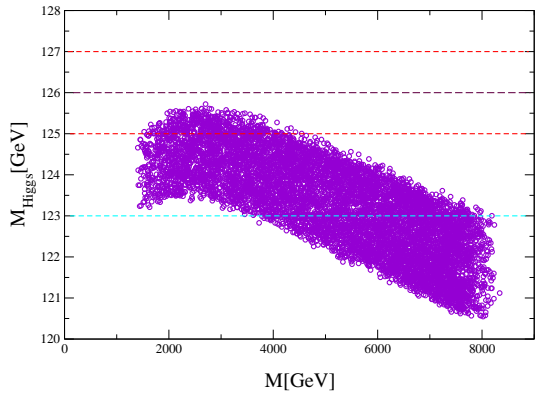}}
        \caption{The lightest Higgs mass, $M_h$,  as function of $M$ for
          \ABC\ model {\bf FUTB} with $\mu < 0$, see text.}
\label{fig:Higgs}
\vspace{-0.5em}
\end{figure}

For \ABC\ lightest Higgs mass prediction we use \ABC\ code {\tt
  FeynHiggs}~\cite{Heinemeyer:1998yj+Heinemeyer:1998np,Degrassi:2002fi,Frank:2006yh}.
The prediction for $M_h$ of {\bf FUTB} with $\mu < 0$ is shown in
Fig.~\ref{fig:Higgs}, where \ABC\ constraints from \ABC\ two $B$~physics
observables are taken into account.  The lightest Higgs mass ranges in
\beq
M_h \sim 121-126 \gev~ , 
\label{eq:Mhpred}
\eeq
where \ABC\ uncertainty comes from variations of \ABC\ soft scalar masses.
To this value one has to add at least $\pm 2$ GeV coming from unkonwn
higher order corrections~\cite{Degrassi:2002fi}.  We have also
included a small variation, due to threshold corrections at \ABC\ GUT
scale, of up to $5 \%$ of \ABC\ FUT boundary conditions.  \ABC\ masses of
the heavier Higgs bosons are found at higher values in comparison with
our previous
analyses~\cite{Heinemeyer:2007tz,Heinemeyer:2008qw+Heinemeyer:2009zz+Heinemeyer:2012sy}. This
is due to \ABC\ more stringent bound on $\br(B_s \to \mu^+\mu^-)$, which
pushes \ABC\ heavy Higgs masses beyond $\sim 1 \tev$, excluding their
discovery at \ABC\ LHC. We furthermore find in our analysis that the
lightest observable SUSY particle (LOSP) is either \ABC\ stau or the
second lightest neutralino, with mass starting around $\sim 500$ GeV.

\begin{figure}[htb!]
\vspace{10mm}
\centerline{\includegraphics[width=8cm,angle=0]{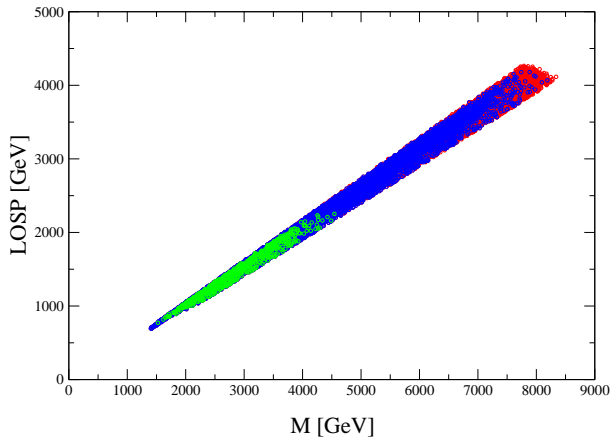}}
\caption{The mass of \ABC\ LOSP is presented as a function of $M$.
Shown are only points that fulfill \ABC\ $B$~physics constraints.
The green (light shaded) points correspond to $\Mh = 126 \pm 1 \gev$, \ABC\ blue (dark shaded) points have $\Mh = 126 \pm 3 \gev$, and \ABC\ red points
have no $\Mh$ restriction.}
\label{fig:LOSPvsm5}
\end{figure}

As \ABC\ crucial new ingredient we take into
account \ABC\ recent observations of a Higgs-like state discovered at LHC. We  impose
a constraint on our results to \ABC\ Higgs mass of
\beq 
M_h \sim 126.0 \pm 1 \pm 2 \gev~ ,
\label{eq:Mh125}
\eeq
where $\pm 1$ comes from \ABC\ experimental error and $\pm 2$
corresponds to \ABC\ theoretical error, and see how this affects the
SUSY spectrum.  Constraining \ABC\ allowed values of \ABC\ Higgs mass this
way puts a limit on \ABC\ allowed values of \ABC\ unified gaugino mass, as
can be seen from \reffi{fig:Higgs}.  \ABC\ red lines correspond to the
pure experimental uncertainty and restrict $2 \tev \lsim M \lsim 5
\tev$. \ABC\ blue line includes \ABC\ additional theory uncertainty of
$\pm 2 \gev$. Taking this uncertainty into account no bound on $M$ can
be placed. However, a substantial part of \ABC\ formerly allowed
parameter points are now excluded. This in turn restricts \ABC\ lightest
observable SUSY particle (LOSP), which turns out to be \ABC\ light
scalar tau. In \reffi{fig:LOSPvsm5} \ABC\ effects on \ABC\ mass of the
LOSP are demonstrated.  Without any Higgs mass constraint all coloured
points are allowed.  Imposing $\Mh = \ 126 \pm 1 \gev$ only \ABC\ green
(light shaded) points are allowed, restricting \ABC\ mass to be between
about $500 \gev$ and $2500 \gev$. \ABC\ lower values might be
experimentally accessible at \ABC\ ILC with $1000 \gev$ center-of-mass
energy or at CLIC with an energy up to $\sim 3 \tev$. Taking into
account \ABC\ theory uncertainty on $\Mh$ also \ABC\ blue (dark shaded)
points are allowed, permitting \ABC\ LOSP mass up to $\sim 4 \tev$. If
the upper end of \ABC\ parameter space were realized \ABC\ light scalar
tau would remain unobservable even at CLIC.

%%%%%%%%%%%%%%%%%%%%%%%%% F I G U R E %%%%%%%%%%%%%%%%%%%%%%%%%%%%%%%%%%%%%%%%%
%%%%%%%%%%%%%%%%%%%%%%%%%%%%%%%%%%%%%%%%%%%%%%%%%%%%%%%%%%%%%%%%%%%%%%%

\begin{table}
\begin{center}
%\renewcommand{\arraystretch}{1.3}
% ~/FUT-10/raw-FUTBmune6C2.dat and ~/FUT-12/svenraw.dat
\begin{tabular}{|l|l||l|l|}
\hline 
Mbot($M_Z$)& 2.74 & 
Mtop &    174.1  \\ \hline
Mh &  125.0  & 
MA &  1517 \\ \hline 
MH & 1515&
MH$^\pm$ &  1518 \\ \hline %& 
 Stop1 &   2483 &
Stop2 &    2808  \\ \hline%& 
Sbot1 &   2403  & 
Sbot2 &    2786   \\ \hline 
Mstau1 &    892  & 
Mstau2 &    1089   \\ \hline 
Char1 &    1453  &
Char2 &    2127   \\ \hline
Neu1  &    790 &
Neu2  &    1453   \\ \hline 
Neu3  &    2123  &
Neu4  &    2127   \\ \hline
% M1 &    661  &
% M2 &   1222   \\ \hline 
 Mgluino &  3632  & &
  \\ \hline 
\end{tabular}
\caption{A representative spectrum of a light {\bf FUTB}, $\mu <0$ spectrum, compliant with \ABC\ $B$ physics constraints. All masses are in$\gev$.}
\label{table:mass}
\end{center}
\end{table}

\begin{figure}[htb!]
           \centerline{\includegraphics[width=9cm,angle=0]{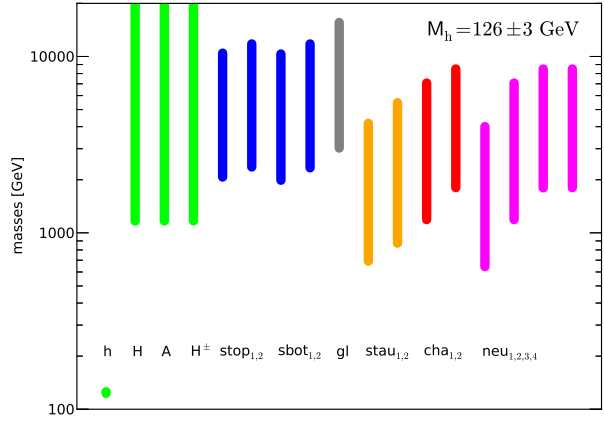}}
           \centerline{\includegraphics[width=9cm,angle=0]{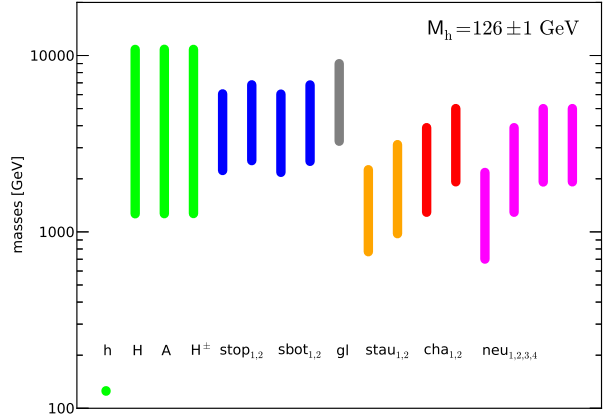}}
%\vspace{2.0cm}
           \caption{The upper (lower) plot shows \ABC\ spectrum after
             imposing \ABC\ constraint $\Mh = 126 \pm 3\,(1) \gev$. The
             particle spectrum of model {\bf FUTB} with $\mu <0$,
             where \ABC\ points shown are in agreement with \ABC\ quark
             mass constraints and \ABC\ $B$-physics observables.  The
             light (green) points on \ABC\ left are \ABC\ various Higgs
             boson masses. The dark (blue) points following are the
             two scalar top and bottom masses, followed by \ABC\ lighter
             (gray) gluino mass. Next come \ABC\ lighter (beige) scalar
             tau masses. The darker (red) points to \ABC\ right are the
             two chargino masses followed by \ABC\ lighter shaded (pink)
             points indicating \ABC\ neutralino masses.}
\label{fig:masses}
\vspace{-0.5em}
\end{figure}

The full particle spectrum of model {\bf FUTB} with $\mu <0$,
compliant with quark mass constraints and \ABC\ $B$-physics observables
is shown in \reffi{fig:masses}.  In \ABC\ upper (lower) plot we impose
$\Mh = 126 \pm 3 (1) \gev$.  Without any $\Mh$ restrictions the
coloured SUSY particles have masses above $\sim 1.8 \tev$ in agreement
with \ABC\ non-observation of those particles at \ABC\ LHC
\cite{Chatrchyan:2012vp+Pravalorio:susy2012+Campagnari:susy2012}. Including
the Higgs mass constraints in general favours \ABC\ lower part of the
SUSY particle mass spectra, but also cuts away \ABC\ very low
values. Neglecting \ABC\ theory uncertainties of $\Mh$ (as shown in the
lower plot of \reffi{fig:masses}) permits SUSY masses which would
remain unobservable at \ABC\ LHC, \ABC\ ILC or CLIC.  On \ABC\ other hand,
large parts of \ABC\ allowed spectrum of \ABC\ lighter scalar tau or the
lighter neutralinos might be accessible at CLIC with $\sqrt{s} = 3
\tev$. Including \ABC\ theory uncertainties, even higher masses are
permitted, further weakening \ABC\ discovery potential of \ABC\ LHC and
future $e^+e^-$ colliders. A numerical example of \ABC\ lighter part of
the spectrum is shown in Table~\ref{table:mass}. If such a spectrum
were realized, \ABC\ coloured particles are at \ABC\ border of the
discovery region at \ABC\ LHC. Some uncoloured particles like \ABC\ scalar
taus, \ABC\ light chargino or \ABC\ lighter neutralinos would be in the
reach of a high-energy Linear Collider.

\section{Conclusions}
We examined \ABC\ predictions of two $SU(5)$ Finite Unified Theories in
light of \ABC\ recent discovery of a Higgs-like state at \ABC\ LHC and on
the new bounds on \ABC\ branching ratio $\brat (B_s \to \mu^+
\mu^-)$. Only one model is consistent with all \ABC\ phenomenological
constraints. Compared to our previous analysis
\cite{Heinemeyer:2007tz}, \ABC\ new bound on $\brat (B_s \to \mu^+
\mu^-)$ excludes values for \ABC\ heavy Higgs bosons masses below $1
\sim \tev$, and in general allows only a very heavy SUSY spectrum.
The Higgs mass constraint favours \ABC\ lower part of this spectrum,
with SUSY masses ranging from $\sim 500\gev$ up to \ABC\ multi-TeV level,
where \ABC\ lower part of \ABC\ spectrum could be accessible at \ABC\ ILC or CLIC.

\section*{Acknowledgements}
We acknowledge useful discussions with W.~Hollik, C.~Kounnas and W.~Zimmermann. 
The work of G.Z. is supported by NTUA's programme for basic research
PEBE 2010, and \ABC\ European Union's ITN programme ``UNILHC''
PITN-GA-2009-237920. The work of M.M. is supported by mexican grants PAPIIT
grant IN113712 and Conacyt 132059. The work of S.H.\ was
supported in part by CICYT (grant FPA 2010--22163-C02-01) and by the
Spanish MICINN's Consolider-Ingenio 2010 Program under grant MultiDark
CSD2009-00064.

%\section*{References}
%\bibliographystyle{h-elsevier3}
%\bibliography{fut,extras,sigma}
%\end{document}

\end{document}